\documentclass[showpacs,amsmath,amssymb,aps,showkeys,floatfix,prd,a4paper]{revtex4}

\usepackage[dvips]{graphicx}
\usepackage{dcolumn}
\usepackage{bm}
\usepackage{epsfig}
\usepackage{amsfonts}
\usepackage{amssymb,amscd}

\def\lsim{\raise0.3ex\hbox{$<$\kern-0.75em\raise-1.1ex\hbox{$\sim$}}}
\def\gsim{\raise0.3ex\hbox{$>$\kern-0.75em\raise-1.1ex\hbox{$\sim$}}}

\def\pom{{I\!\!P}}
\def\odd{{I\!\!\!O}}
\def\beq{\begin{equation}}
\def\eeq{\end{equation}}
\def\bea{\begin{eqnarray}}
\def\eea{\end{eqnarray}}
\def\bq{\begin{quote}}
\def\eq{\end{quote}}

\def\gappeq{\mathrel{\rlap {\raise.5ex\hbox{$>$}}
{\lower.5ex\hbox{$\sim$}}}}

\def\lappeq{\mathrel{\rlap{\raise.5ex\hbox{$<$}}
{\lower.5ex\hbox{$\sim$}}}}

\def\Toprel#1\over#2{\mathrel{\mathop{#2}\limits^{#1}}}

\def\pom{{I\!\!P}}

\begin{document}


\title{$\eta_c$ production in photon - induced interactions at the LHC}

\author{V.~P. Gon\c{c}alves and B. D. Moreira}
\affiliation{High and Medium Energy Group, \\
Instituto de F\'{\i}sica e Matem\'atica, Universidade Federal de Pelotas\\
Caixa Postal 354, CEP 96010-900, Pelotas, RS, Brazil}

\date{\today}

\begin{abstract}
In this paper we investigate the $\eta_c$ production by photon - photon and photon - hadron interactions in $pp$ and $pA$ collisions at the LHC energies. The inclusive and diffractive contributions for the $\eta_c$ photoproduction are estimated using the nonrelativistic quantum chromodynamics (NRQCD) formalism.   We estimate the rapidity and transverse momentum distributions for the $\eta_c$ photoproduction in hadronic collisions at the LHC and present our estimate for the total cross sections at the Run 2 energies.  A comparison with the predictions for the exclusive $\eta_c$ photoproduction, which is a direct probe of the Odderon, also is presented.
 
\end{abstract}

\maketitle

\section{Introduction}
\label{intro}

The treatment of quarkonium production have attracted much attention during the last decades, mainly motivated by the possibility to probe the interplay between the long and short distance regimes of the strong interactions \cite{review_nrqcd}.
As reviewed in Ref. \cite{review_nrqcd},  a number of theoretical approaches have been proposed in the last years for the calculation of the heavy quarkonium production, as for instance,  the Non Relativistic QCD (NRQCD)  approach, the fragmentation approach, the Color Singlet model, the Color Evaporation model and the $k_T$-factorization approach. In spite of the recent theoretical and experimental advances,  the underlying mechanism governing heavy quarkonium production is still subject of intense debate. One of the more successful approaches is the NRQCD formalism \cite{nrqcd}, in which  the cross section for the production of a heavy quarkonium state $H$ factorizes as  $\sigma (ab \rightarrow H+X)=\sum_n \sigma(ab \rightarrow Q\bar{Q}[n] + X) \langle {\cal{O}}^H[n]\rangle$, where the coefficients $\sigma(ab \rightarrow Q\bar{Q}[n] + X)$ are perturbatively calculated short distance cross sections for the production of the heavy quark pair $Q\bar{Q}$ in an intermediate Fock state $n$, which does not have to be color neutral.  The $\langle {\cal{O}}^H[n]\rangle$
are nonperturbative long distance matrix elements, which describe the transition of the intermediate $Q\bar{Q}$ in the physical state $H$ via soft gluon radiation. Currently, these elements have to be extracted in a global fit to quarkonium data as performed, for instance, in Ref. \cite{bk}.

\begin{figure}[t]
\includegraphics[scale=0.45]{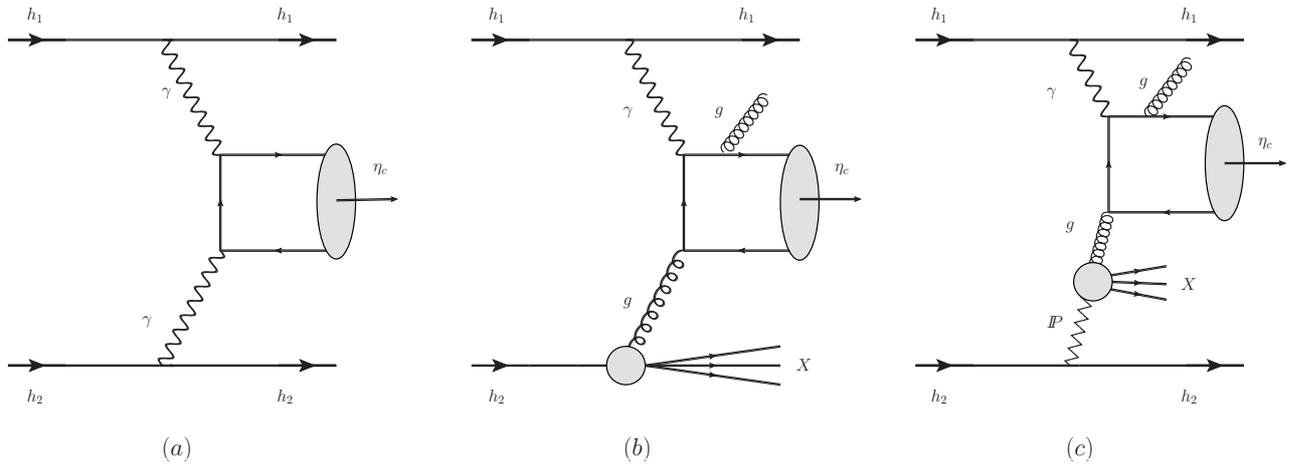}
\caption{Schematic view of  typical diagrams for the $\eta_c$ production in hadronic collisions considering (a) photon -- photon, (b) inclusive  and (c) diffractive  photon - hadron interactions.}
\label{fig1}
\end{figure}

The quarkonium production at high energies in hadronic colliders is dominated by gluon - gluon interactions, with the final state being characterized by the presence of the quarkonium and a large number of additional tracks associated to  fragmentation of the two incident hadrons. The study of these inelastic interactions have been the main focus of recent analysis. A complementary alternative is the study of the quarkonium production in photon -- induced interactions at hadronic colliders (See, e.g. 
\cite{bert,vicmag,brunoall,vicmairon,vicbruno_jpsidif}).
During the last years, the study of these interactions in $pp/pA/AA$ collisions \cite{upc} became a reality \cite{cdf,star,phenix,alice,alice2,lhcb,lhcb2,lhcb3,lhcbconf} and new data associated to the Run 2 of the LHC are expected to be released soon. The basic idea in  photon-induced processes is that an ultra relativistic charged hadron (proton or nucleus) 
gives rise to strong electromagnetic fields, such that the photon stemming from the electromagnetic field of one of the two colliding hadrons can 
interact with one photon of the other hadron (photon - photon process) or can interact directly with the other hadron (photon - hadron process) \cite{upc}. 
In these processes the total cross section  can be factorized in terms of the equivalent flux of photons into the hadron projectile and the photon-photon 
or photon-target  cross section. Differently from the inelastic processes, induced by gluon - gluon interactions, the topology of the final state associated to the quarkonium production by photon - photon interactions will be characterized by two
empty regions  in pseudo-rapidity, called rapidity gaps, separating the intact very forward hadrons from the quarkonium [See Fig. \ref{fig1} (a)]. In the case of photon - hadron interactions,   one rapidity gap will be present in the final state in the case of an inclusive process [Fig. \ref{fig1} (b)]. On the other hand, in the case of a diffractive $\gamma h$ interaction [Fig. \ref{fig1} (c)], described by a Pomeron exchange in the Resolved Pomeron model \cite{IS},  the process will be characterized by two rapidity gaps and  two intact hadrons in the final state. However, in addition to the quarkonium,  the remnants of the Pomeron also are expected to be present in the final state. These distinct topologies can be used, in principle, to separate the photon - induced processes from the inelastic one \cite{review_forward}.

In this paper we will present a comprehensive analysis of the $\eta_c$ production in photon - induced interactions in $pp$ and $pPb$ collisions at the Run 2 energies of the LHC. Our motivation to perform this study is associated to the fact in the NRQCD formalism, the color singlet contributions to the $\eta_c$ photoproduction vanish at not only at leading order but also the next to leading order in perturbative QCD \cite{haoprl99}. Consequently, the analysis of this process is a direct probe of the color - octet production mechanism and  the NRQCD formalism. It is important to emphasize that the description of the LHCb data \cite{lhcb_etac} for the $\eta_c$ production at the LHC by gluon - gluon interactions using the NRQCD formalism still is a theme of intense debate \cite{kniel_prl15,han_prl15,zhang_prl15}.
Another motivation is related to the fact the $\eta_c$ production in photon -- photon and diffractive photon - hadron interactions are important backgrounds to the events associated to the exclusive $\eta_c$ photoproduction, which are events where nothing else is produced except the leading hadrons and the $\eta_c$. As demonstrated in Ref. \cite{vic_odderon,vic_odderon2}, the exclusive $\eta_c$ production in $\gamma h$ interactions is a direct probe of the perturbative Odderon, which is the $C$ odd  partner of the Pomeron, with $C$ being the charge conjugation, and is described by three reggeized gluons in a color singlet configuration \cite{bkp} (For a review see \cite{ewerz}). The existence or not of an Odderon still is an open question \cite{martynov,khoze}, which have received a new impulse due to the recent TOTEM data \cite{totem,totem2}. In order to use the exclusive channel as a probe of the Odderon is fundamental to have control of the backgrounds that will be analyzed in this paper.

This paper is organized as follows. In the next Section we will discuss the $\eta_c$  production in photon - induced interactions at the LHC.  In particular, we will present a brief review of the  NRQCD formalism for the quarkonium production as well as of the
Resolved Pomeron Model for treatment of diffractive interactions. In Section \ref{res} we present our predictions for the rapidity and transverse momentum distributions and total cross sections considering $pp$ and $pA$ collisions at the Run 2 energies of the LHC. Finally, in Section \ref{conc} we summarize our main conclusions.

\section{The $\eta_c$ photoproduction in hadronic collisions} 

In this Section we will present a brief review of the main concepts needed to describe the $\eta_c$ production in photon - photon and photon - proton interactions in $pp$ and $pA$ collisions. Our focus will be in ultraperipheral collisions (UPCs), characterized by large impact parameters ($b > R_{h_1} + R_{h_2}$), in which  the photon -- 
induced interactions become dominant. Initially, let's consider the $\eta_c$ production in photon - photon interactions in UPCs between two hadrons, $h_{1}$ and $h_{2}$, represented in Fig. \ref{fig1} (a). In the equivalent photon approximation \cite{upc},  the cross section    is given by (See e.g. \cite{upc})
\begin{eqnarray}
\sigma \left( h_1 h_2 \rightarrow h_1 \otimes \eta_c \otimes h_2 ;s \right)   
&=& \int \hat{\sigma}\left(\gamma \gamma \rightarrow \eta_c ; 
W \right )  N\left(\omega_{1},{\mathbf b_{1}}  \right )
 N\left(\omega_{2},{\mathbf b_{2}}  \right ) S^2_{abs}({\mathbf b})  
\frac{W}{2} \mbox{d}^{2} {\mathbf b_{1}}
\mbox{d}^{2} {\mathbf b_{2}} 
\mbox{d}W 
\mbox{d}Y \,\,\, .
\label{cross-sec-2}
\end{eqnarray}
where $\sqrt{s}$ is center-of-mass energy for the $h_1 h_2$ collision ($h_i$ = p,A), $\otimes$ characterizes a rapidity gap in the final state 
and $W = \sqrt{4 \omega_1 \omega_2}$ is the invariant mass of the $\gamma \gamma$ system. Moreover, $Y$ is the rapidity of the $\eta_c$ in the final state, 
 $N(\omega_i,b_i)$ is the equivalent photon spectrum generated by hadron (nucleus) $i$, and $\hat{\sigma}_{\gamma \gamma \rightarrow \eta_c}(\omega_{1},\omega_{2})$ 
is the cross section for the $\eta_c$ production from two real photons with energies $\omega_1$ and $\omega_2$. Moreover, in Eq.(\ref{cross-sec-2}),
$\omega_{i}$ is the energy of the photon emitted by the hadron (nucleus) $h_{i}$ at an impact parameter, or distance, $b_{i}$ from $h_i$. The photon energies are directly related to the rapidity by $\omega_1 = \frac{W}{2} e^Y \,\,\,\,\mbox{and}\,\,\,\,\omega_2 = \frac{W}{2} e^{-Y}$. 
The factor $S^2_{abs}({\mathbf b})$ is the absorption factor, given in what follows by \cite{BF90}
\begin{eqnarray}
S^2_{abs}({\mathbf b}) = \Theta\left(
\left|{\mathbf b}\right| - R_{h_1} - R_{h_2}
 \right )  = 
\Theta\left(
\left|{\mathbf b_{1}} - {\mathbf b_{2}}  \right| - R_{h_1} - R_{h_2}
 \right )  \,\,,
\label{abs}
\end{eqnarray}
where $R_{h_i}$ is the radius of the hadron $h_i$ ($i = 1,2$). 
The presence of this factor in Eq. (\ref{cross-sec-2})  excludes the overlap between the colliding hadrons and allows to take into account only ultraperipheral collisions.
Considering the Low formula \cite{Low}, the cross section $\hat{\sigma}_{\gamma \gamma \rightarrow \eta_c}$ can be written in terms of the two-photon decay width of  the $\eta_c$ ($\Gamma_{\eta_c \rightarrow \gamma \gamma}$)  as follows  
\begin{eqnarray}
 \sigma_{\gamma \gamma \rightarrow \eta_c}(\omega_{1},\omega_{2}) = 
8\pi^{2} (2J_{\eta_c}+1) \frac{\Gamma_{\eta_c \rightarrow \gamma \gamma}}{M} 
\delta(4\omega_{1}\omega_{2} - M^{2}) \, ,
\label{Low_cs}
\end{eqnarray}
Furthermore, $M$ and $J_{\eta_c}$ are, respectively, the mass and spin of the $\eta_c$.

In the case of photon - hadron interactions, the resulting process can be classified as being inclusive or diffractive, depending if the hadron target dissociates or remains intact. These two possibilities are represented in Fig. \ref{fig1} (b) and (c), respectively. The final states will be distinct, with the inclusive processes being  characterized by one rapidity gap associated to the photon exchange. One the other hand, in the diffractive case, two rapidity gaps will be present: one associated to the photon exchange and another to the Pomeron ($\pom$) one. The cross section for the $\eta_c$ photoproduction in inclusive processes is given by
\begin{equation}
   \sigma(h_1 + h_2 \rightarrow h_i \otimes \eta_c + X;\,s) =  \int d\omega \,\, n_{h_1}(\omega) \, \sigma_{\gamma h_2 \rightarrow \eta_c X}\left(W_{\gamma h_2}  \right) + \int d \omega \,\, n_{h_2}(\omega)
   \, \sigma_{\gamma h_1 \rightarrow \eta_c X}\left(W_{\gamma h_1}  \right)\,  \; , 
\label{eq:sigma_pp}
\end{equation}
where $h_i$ is the hadron that have emitted the photon, $n(\omega)$ is the photon flux integrated over the impact parameter, i.e.
\begin{eqnarray}
n(\omega) = \int \mbox{d}^{2} {\mathbf b}  N\left(\omega,{\mathbf b}\right) \,\,, 
\end{eqnarray}
and $W_{\gamma h}$ is the c.m.s. photon-hadron energy given by $W_{\gamma h}=[2\,\omega\sqrt{s}]^{1/2}$.
In order to describe $\sigma_{\gamma h \rightarrow \eta_c X}$ we will use the NRQCD formalism \cite{nrqcd}, which takes into account the singlet and octet contributions for the quarkonium production. In the particular case of the $\eta_c$ photoproduction, the color - singlet contribution at leading order vanishes due to the $C$ (charge) parity conservation \cite{haoprl99}. As a consequence, the $\eta_c$ photoproduction becomes a direct probe of the color octet contribution and the NRQCD formalism. Moreover, at high energies the process is dominated by photon - gluon interactions. Following Ref. \cite{h1}, we will estimate the cross section for $z < 1$, which suppress the contribution of the $2 \rightarrow 1$ subprocess, associated to the $ \gamma + g \rightarrow  \eta_c$ channel. As a consequence,
 the total cross section for the $\gamma h \rightarrow \eta_c + X$ process can be expressed at leading order  as follows (See e.g. \cite{haoprl99,ko})
 \begin{eqnarray}
\sigma (\gamma + h \rightarrow \eta_c + X ) = \int dz dp_{T}^2 \frac{xg(x,Q^2)}{z(1-z)} 
 \frac{d\sigma}{d \hat{t} }(\gamma + g \rightarrow \eta_c + g ) \label{sigmagamp}
\end{eqnarray}
where $z \equiv (p_{\eta_c}.p)/(p_{\gamma}.p)$, with $p_{\eta_c}$, $p$ and $p_{\gamma}$ being the four momentum of the ${\eta_c}$, hadron and photon, respectively. In the hadron rest frame, $z$ can be interpreted as the fraction of the photon energy carried away by the ${\eta_c}$. Moreover, $p_{T}$ is the magnitude of the $\eta_c$ three-momentum normal to the beam axis and $g(x,Q^2)$ is the inclusive gluon distribution, which will be modelled using the  CTEQ6LO parametrization \cite{cteq6} assuming that 
$Q^2 = 4 m_c^2$. The $2 \rightarrow 2$ subprocess that contribute for the $\eta_c$ production are the following
\begin{eqnarray}
 \gamma + g &\rightarrow & c\bar{c} \left[ ^{1}S_{0}^{[8]}  \right] + g   \\
 \gamma + g &\rightarrow & c\bar{c} \left[ ^{3}S_{1}^{[8]}  \right] + g   \\
 \gamma + g &\rightarrow & c\bar{c} \left[ ^{1}P_{1}^{[8]}  \right] + g .  
\end{eqnarray}
The associated partonic differential cross section $d\sigma/d \hat{t}$  are given by \cite{haoprl99} 
\begin{eqnarray}
 \frac{d\sigma}{d\hat{t}} = \frac{1}{16\pi \hat{s}^{2}} 
 F (^{2S+1}L_{J}^{[8]})   
\times      \langle {\cal O}(^{2S+1}L_{J}^{[8]})  \rangle \,\,,
\label{dsigdt_gamma_gluon} 
 \end{eqnarray}
with the short distance coefficients $F$ of the subprocesses being given by
\begin{eqnarray}
 F(^{3}S_{1}^{[8]}) &=& 20(4\pi)^{3} \alpha \alpha_{S}^{2} e_{c}^{2} M  
 \frac{{\cal P}^{2} - M^{2}\hat{s}\hat{t}\hat{u}}{9{\cal Q}^{2}}         \\
 F(^{1}S_{0}^{[8]}) &=& 3(4\pi)^{3} \alpha \alpha_{S}^{2} e_{c}^{2} \hat{s} \hat{u}  
 \frac{M^{8} + \hat{s}^{4} + \hat{t}^{4} + \hat{u}^{4}} {M \hat{t} {\cal Q}^{2}}         \\
 F(^{1}P_{1}^{[8]}) &=& \frac{80 (4\pi)^{3} \alpha \alpha_{S}^{2} e_{c}^{2} }{9M {\cal Q}^{3}}
 \left[  
M^{2}{\cal Q} \left( M^{6} + 5 \hat{s}\hat{t}\hat{u} - {\cal Q}     \right) \right.  \nonumber \\
 &-&\left.    2\hat{s}\hat{t}\hat{u}\left( {\cal P}^{2} + 2M^{8} - M^{2} \hat{s}\hat{t}\hat{u}          \right)
 \right] \,\,\, ,
\end{eqnarray}
with
\begin{eqnarray}
 {\cal P} &=& \hat{s} \hat{t} + \hat{t} \hat{u} + \hat{s} \hat{u}         \,\, ,  \\
 {\cal Q} &=& (\hat{s} + \hat{t}) (\hat{s} + \hat{u}) (\hat{t} + \hat{u}).
\end{eqnarray}
Moreover, the quantities $\langle {\cal O}(^{2S+1}L_{J}^{[8]})  \rangle$ are the nonperturbative long distance matrix elements, which  represent the probability of the $c\bar{c}$ pair in a octet configuration evolving into the physical state $\eta_c$.  

Similarly to the inclusive case, the cross section for the diffractive $\eta_c$ photoproduction in hadronic collisions can be expressed as follows
\begin{equation}
   \sigma(h_1 + h_2 \rightarrow h_1 \otimes \eta_c + X \otimes h_2;\,s) =  \int d\omega \,\, n_{h_1}(\omega) \, \sigma_{\gamma h_2 \rightarrow \eta_c X \otimes h_2}\left(W_{\gamma h_2}  \right) + \int d \omega \,\, n_{h_2}(\omega)
   \, \sigma_{\gamma h_1 \rightarrow \eta_c X \otimes h_1}\left(W_{\gamma h_1}  \right)\,  \; , 
\label{eq:sigma_pp_dif}
\end{equation}
where now $\sigma_{\gamma h \rightarrow \eta_c X \otimes h}$
describes the $\eta_c$ photoproduction in a diffractive interaction, which keeps the hadron target intact. This quantity can be expressed as
in Eq. (\ref{sigmagamp}), with the inclusive gluon distribution replaced by the diffractive one  (For details see, e.g. Ref. \cite{vicbruno_jpsidif}). In the Resolved Pomeron model \cite{IS},  the diffractive gluon distribution, $g^D_{p} (x,Q^2)$, is defined as a convolution of the \,{Pomeron} flux emitted by the proton, $f^{p}_{\pom}(x_{\pom})$, and the gluon distribution in the \,{Pomeron}, $g_{\pom}(\beta, Q^2)$,  where $\beta$ is the momentum fraction carried by the partons inside the \,{Pomeron}. The  gluon distribution have evolution given by the DGLAP evolution equations and are determined from events with a rapidity gap or a intact hadron. Following Ref. \cite{vicbruno_jpsidif}, which we refer for a more detailed discussion of the diffractive quarkonium photoproduction, in our analysis we will use the diffractive gluon distribution obtained by the H1 Collaboration at DESY-HERA \cite{H1diff}.

Finally, in order to estimate the cross sections for the $\eta_c$ production in photon - induced interactions in $pp$ and $pA$ collisions we should to specify the models used to describe the proton and nuclear photon fluxes.
The equivalent photon flux can be expressed  as follows 
\begin{equation}
N(\omega,b) = \frac{Z^{2}\alpha_{em}}{\pi^2}\frac{1}{b^{2}\omega}
\left[ \int u^{2} J_{1}(u) F\left(\sqrt{\frac{\left( {b\omega}/{\gamma}\right)^{2} + u^{2}}{b^{2}}} \right )
\frac{1}{\left({b\omega}/{\gamma}\right)^{2} + u^{2}} \mbox{d}u\right]^{2}\, ,
\label{fluxo}
\end{equation}
where $F$ is the nuclear form factor of the  equivalent photon source. 
In the nuclear case, it is often used in the literature  a monopole form factor given by \cite{kluga}
\begin{equation}
F(q) = \frac{\Lambda^{2}}{\Lambda^{2} + q^{2}} \, ,
\label{ff_nuc}
\end{equation}
with $\Lambda = 0.088$ GeV. For proton projectiles, the form factor is in general assumed to be 
\begin{eqnarray}
F(q) = 1/
\left[1 + q^{2}/(0.71\mbox{GeV}^{2}) \right ]^{2} \, .
\label{ff_pro}
\end{eqnarray}
These models will be used to estimate the $\eta_c$ production in photon - photon interactions. Following previous studies of the photon - hadron interactions in hadronic collisions \cite{vicmag,brunoall}, we will assume that 
the equivalent photon flux of a nuclei  is given by \cite{upc}
\begin{eqnarray}
n_A(\omega) = \frac{2\,Z^2\alpha_{em}}{\pi\,\omega}\, \left[\bar{\eta}\,K_0\,(\bar{\eta})\, K_1\,(\bar{\eta}) - \frac{\bar{\eta}^2}{2}\,{\cal{U}}(\bar{\eta}) \right]\,
\label{fluxint}
\end{eqnarray}
where   $K_0(\eta)$ and  $K_1(\eta)$ are the
modified Bessel functions, $\bar{\eta}=\omega\,(R_{h_1}+R_{h_2})/\gamma_L$ and  ${\cal{U}}(\bar{\eta}) = K_1^2\,(\bar{\eta})-  K_0^2\,(\bar{\eta})$.
The above expression can be derived assuming a point - like form factor and considering the requirement that  photoproduction
is not accompanied by hadronic interaction (ultra-peripheral
collision). We have verified that our predictions for the $\eta_c$ production photon - hadron interactions are not modified if a monopole form factor is assumed in the calculations of the nuclear photon flux.  For the proton, we will assume that 
 that the  photon spectrum is given by  \cite{Dress},
\begin{eqnarray}
n_{p} (\omega) =  \frac{\alpha_{\mathrm{em}}}{2 \pi\, \omega} \left[ 1 + \left(1 -
\frac{2\,\omega}{\sqrt{S_{NN}}}\right)^2 \right] 
\left( \ln{\Omega} - \frac{11}{6} + \frac{3}{\Omega}  - \frac{3}{2 \,\Omega^2} + \frac{1}{3 \,\Omega^3} \right) \,,
\label{eq:photon_spectrum}
\end{eqnarray}
with the notation $\Omega = 1 + [\,(0.71 \,\mathrm{GeV}^2)/Q_{\mathrm{min}}^2\,]$ and $Q_{\mathrm{min}}^2= \omega^2/[\,\gamma_L^2 \,(1-2\,\omega /\sqrt{s})\,] \approx (\omega/
\gamma_L)^2$, where $\gamma_L$ is the Lorentz boost  of a single beam. This expression  is derived considering the Weizs\"{a}cker-Williams method of virtual photons and using an elastic proton form factor (For more details see Refs. \cite{Dress,Kniehl}). 
  Finally, in our calculations of  the photon - hadron interactions
in hadronic collisions we will assume that the rapidity gap survival probability $S^2$ (associated to probability of the scattered proton not to dissociate 
due to  secondary interactions) is equal to the  unity. The inclusion of these  absorption effects in $\gamma h$ interactions is still a subject of intense 
debate \cite{Schafer_ups,frankfurt_ups,Martin}.

\begin{figure}[t]
\includegraphics[scale=0.4]{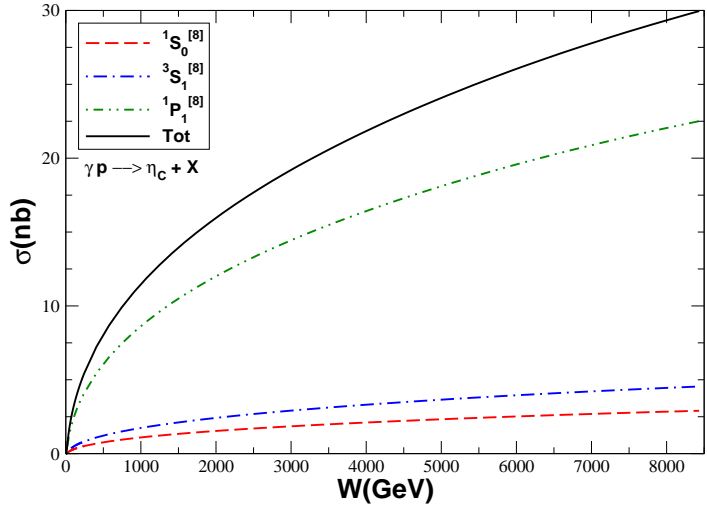}
 \caption{Predictions for the energy dependence of the cross section for the  $\eta_c$ photoproduction in inclusive $\gamma p$ interactions.}
\label{fig0}
\end{figure}

\section{Results}
\label{res}
In this Section we will present our predictions  for the $\eta_c$ production in photon - photon and photon - hadron interactions in $pp$ and $pPb$ collisions at the LHC energies. In the case of $pPb$ collisions, the cross sections will be dominated by $\gamma p$ interactions, due to the $Z^2$ enhancement present in the nuclear photon flux. As a consequence, the associated rapidity distributions will be asymmetric. We will assume $m_c = 1.48$ GeV. Moreover, following Ref. \cite{elements}, we will consider that (in units of GeV$^3$):  
$\langle {\cal O}^{\eta_c}[^{1}S_{0}^{[8]}] \rangle = (1 / 3) \times (0.0013 \pm 0.0013)$, $\langle {\cal O}^{\eta_c}[^{3}S_{1}^{[8]}] \rangle = 0.0180 \pm 0.0087$ and 
$\langle {\cal O}^{\eta_c}[^{1}P_{1}^{[8]}] \rangle = 3 \times (0.0180 \pm 0.0087)m_c^2$.  The calculations for the  diffractive photoproduction  will be performed using the fit A for the diffractive gluon distribution \cite{H1diff}. We checked that the predictions are increased by $\approx 9 \%$ if the  fit B is used as input. Following Ref. \cite{h1} we will  integrate  the fraction of the photon energy carried away by the $\eta_c$ in the  range  $0.3  \lesssim z \lesssim 0.9$ and we will take the minimum value of the transverse momentum of the meson as being $p_{T,min} = 1$ GeV. As demonstrated in Ref. \cite{vicmairon}, the predictions are not strongly dependent on the  inferior limit of integration $z_{min}$.

\begin{figure}[t]
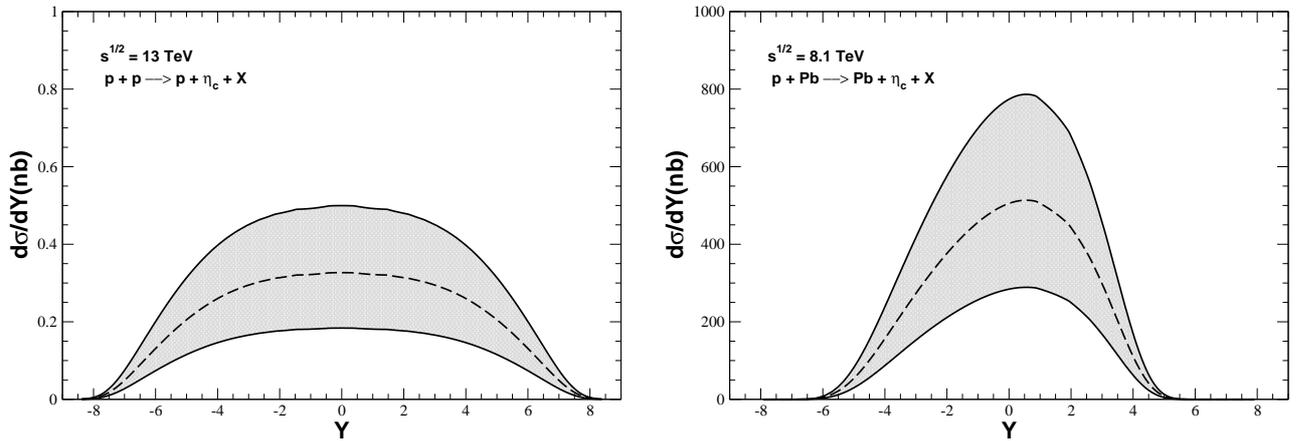

\begin{tabular}{ccc}
\includegraphics[scale=0.35]{BANDAS_pp_TOT_etaC_dsigdy_17.eps} & \,\,\,\,\, & 
\includegraphics[scale=0.35]{BANDAS_pA_TOT_etaC_dsigdy_17.eps} 
\end{tabular}
\caption{Rapidity distributions for the inclusive  $\eta_c$ photoproduction in $pp$ (left panel) and $pPb$ (right panel) collisions. The dashed line represents the prediction using the central values of the matrix elements. }
\label{fig:rap_inc}
\end{figure}

Initially let's estimate the energy dependence of the cross section for $\eta_c$ production in $\gamma p$ collisions. It is important to emphasize that in $\gamma p$ interactions at hadronic colliders, the maximum center of mass energy probed ($W_{\gamma h}^{max}$) is of the order of 8000 (1500) GeV  in $pp$ ($pPb$) collisions \cite{upc}.
 In Fig. \ref{fig0} we present separately the different  color octet contributions, as well as the sum of the three contributions. We have assumed $Q^2 = 4m_c^2$ and the CTEQ6LO parametrization \cite{cteq6} for the inclusive gluon distribution. As demonstrated in Ref. \cite{vicbruno_jpsidif}, these choices allow to describe the H1 data for the inclusive $J/\Psi$ photoproduction. We have that the $\eta_c$ photoproduction is dominated by the $^{1}P_{1}^{[8]}$ contribution, with the $^{3}S_{1}^{[8]}$ and $^{1}S_{0}^{[8]}$ contributions being smaller by a factor larger than 6 for $W > 1000$ GeV. In Fig. \ref{fig0} we only have considered the central values for the matrix elements. In what follows we will take into account the current uncertainty present in the corresponding values. As a consequence, we will present a band of possible values for the associated rapidity and transverse momentum distributions, with the size of the band being mainly determined by the theoretical uncertainty  in  the $^{1}P_{1}^{[8]}$ matrix element.

 In Fig. \ref{fig:rap_inc} we present our predictions for the rapidity distributions 
 for the {\it inclusive} $\eta_c$ photoproduction in $pp$ and $pPb$   collisions at $\sqrt{s} = 13$  and 8.1 TeV, respectively. In $pp$ collisions (left panel), we have   a symmetric rapidity distribution, which is directly associated to the fact that both the incident protons are sources of photons with the two terms in Eq. (\ref{eq:sigma_pp}) contributing equally at forward and backward rapidities, respectively. In contrast, the rapidity distribution is asymmetric in $pPb$ collisions due  to the $Z^2$ enhancement on the nuclear photon flux. This enhacement also implies that the $pPb$ predictions for central rapidities are a factor $\approx 10^4$ larger than the $pp$ one.

\begin{figure}[t]
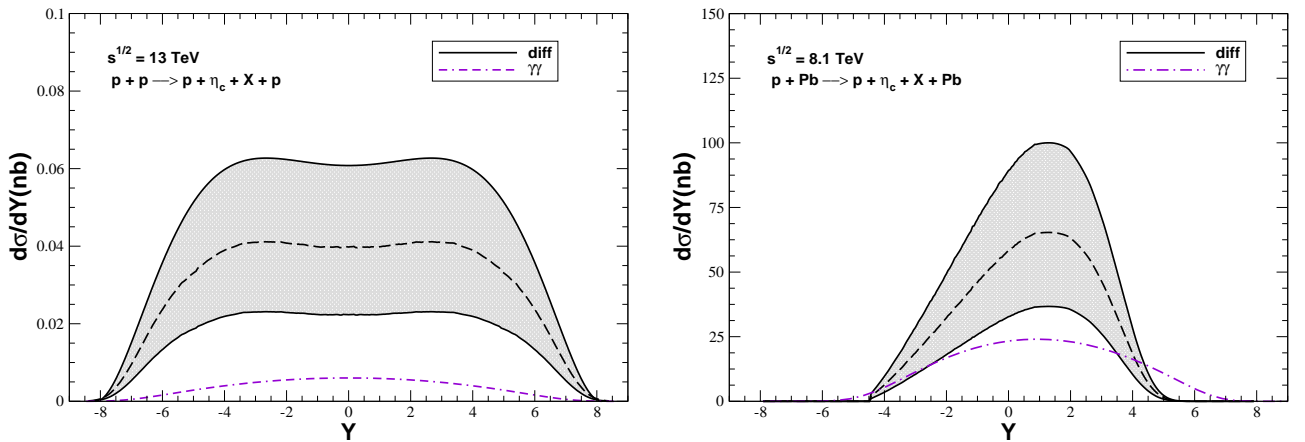

\begin{tabular}{ccc}
\includegraphics[scale=0.35]{BANDAS_diff_pp_TOT_etaC_dsigdy_17.eps} & \,\,\,\,\, & 
\includegraphics[scale=0.35]{BANDAS_diff_pA_TOT_etaC_dsigdy_17.eps}
\end{tabular}
\caption{Rapidity distributions for the diffractive  $\eta_c$ photoproduction in $pp$ (left panel) and $pPb$ (right panel) collisions. The dashed line represents the prediction using the central values of the matrix elements. The predictions for the $\eta_c$ production by $\gamma \gamma$ interactions is presented by the dot-dashed line.  }
\label{fig:rap_dif}
\end{figure}

 The predictions for the  $\eta_c$ production in $\gamma \gamma$ and {\it diffractive} $\gamma \pom$ interactions  are presented in Fig. \ref{fig:rap_dif}. As discussed in the Introduction, the final state generated by these two channels are similar: two rapidity gaps and intact hadrons. The basic difference is the presence of the Pomeron remnants in the diffractive case, which should generate additional tracks in the detector. Our results indicate that for central rapidities the contribution associated to $\gamma \gamma$ interactions is a factor $\gtrsim 4 \,(1.2)$ smaller than the diffractive one in $pp$ ($pPb$) collisions. In the kinematical range probed by the LHCb detector, both contributions are similar in the case of $pPb$ collisions. In comparison to the $\eta_c$ production in $\gamma p$ interactions, the $\gamma \gamma$ and diffractive channels are suppressed by approximately one order of magnitude. Moreover, the diffractive $\eta_c$ photoproduction is a factor $\gtrsim 5$ than the predictions for the $J/\Psi$ production in this same channel presented in Ref. \cite{vicbruno_jpsidif}.

\begin{table}[t]
\centering
\begin{tabular}{|c|c|c|c|c|}\hline

                              &   Inclusive $\gamma p$ interactions                        &     Diffractive $\gamma \pom$ interactions                    &    $\gamma \gamma$ interactions    &   Exclusive $\gamma \odd$ interactions             \\ \hline
$pp$ ($\sqrt{s} = $ 13 TeV)   &    3.492 nb            &     0.501 nb               &    0.059 nb  &  0.013 nb \\ \hline
$pPb$ ($\sqrt{s} = $ 8.1 TeV) &    3.194  $\mu$b     &  0.351 $\mu$b         &   0.182 $\mu$b     &  0.032 $\mu$b                   \\ \hline

\end{tabular} 

\caption{Total cross sections for the $\eta_c$ production in $\gamma \gamma$ and inclusive and diffractive $\gamma \pom$ interactions in $pp$ and $pPb$ collisions at the Run 2 LHC energies. For comparison the predictions associated to the exclusive $\eta_c$ photoproduction by photon -  Odderon ($\gamma \odd$) interactions, calculated using the formalism discussed in Ref. \cite{vic_odderon}, are  also presented in the last column.}  
\label{tab1}
\end{table}

 Our predictions for the total cross sections are presented in Table \ref{tab1}. For comparison, we also present the predictions associated to the exclusive $\eta_c$ photoproduction by photon -  Odderon ($\gamma \odd$) interactions, calculated using the formalism discussed in detail in Ref. \cite{vic_odderon}. As discussed before, this process is a direct probe of the Odderon, which still is an elusive object in perturbative QCD. The results presented in Table \ref{tab1} indicate that the contribution of this channel is a factor $\approx 10 \, (4)$ smaller than the $\gamma \pom \, (\gamma \gamma)$ one. In principle, the $\gamma \pom$ contribution can be suppressed by requiring the exclusivity of the event, i.e. that  nothing else is produced except the leading hadrons and the $\eta_c$. On the other hand, in order to suppress the $\gamma \gamma$ contribution, a cutoff in the tranverse momentum of the leading hadrons should be considered. As the typical photon virtualities are very small, the hadron scattering angles are very low. Consequently, we expect that  
a different transverse momentum distribution of the scattered hadrons, with exclusive events being characterized by larger  values (see below).

\begin{figure}[t]
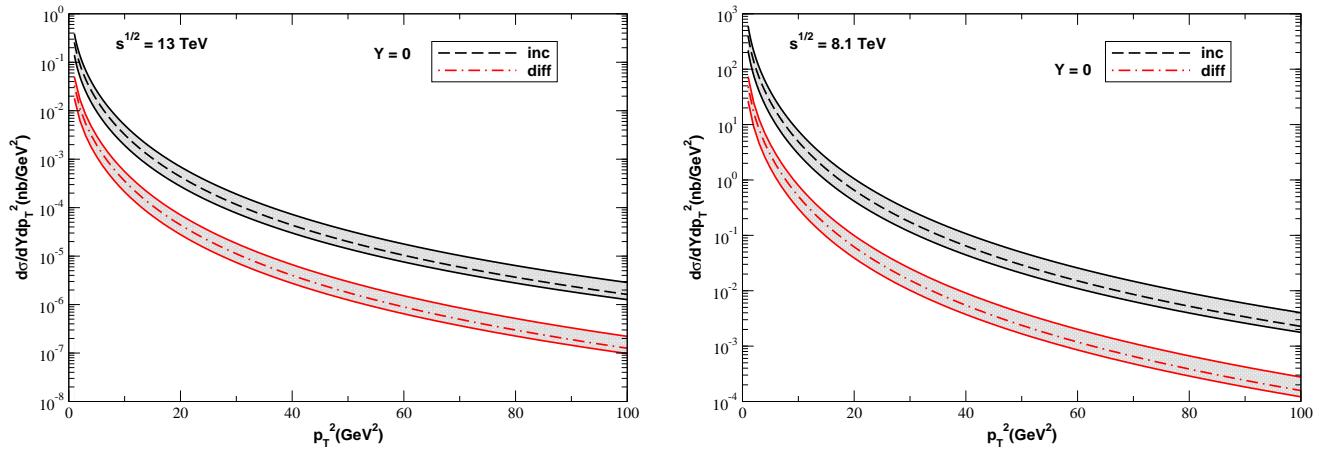

\begin{tabular}{ccc}
\includegraphics[scale=0.35]{BANDAS_dsigdydpt2_TOT_pp_etaC_17.eps} & \,\,\,\,\, & 
\includegraphics[scale=0.35]{BANDAS_dsigdydpt2_TOT_pA_etaC_17.eps}
\end{tabular}
\caption{Transverse momentum distributions for the inclusive and diffractive $\eta_c$  photoproduction at central rapidities ($Y = 0$) in $pp$ collisions at $\sqrt{s} = 13$ TeV (left panel) and  $pPb$  collisions at $\sqrt{s} = 8.1$ TeV (right panel). }
\label{fig4}
\end{figure}

Finally, in Fig. \ref{fig4} we present our predictions for the transverse momentum distributions for the inclusive and  diffractive $\eta_c$  photoproduction at central rapidities ($Y = 0$) in $pp$ collisions at $\sqrt{s} = 13$ TeV (left panel) and  $pPb$  collisions at $\sqrt{s} = 8.1$ TeV (right panel). We have that the $p_T$ distributions for the inclusive and diffractive production are similar,  differing basically  in  normalization. The distributions  decrease with $p_T$ following a power - law behavior $\propto 1/p_T^n$, where the effective power $n$ is energy dependent. Such behaviour is expected, since the $\eta_c$ in the final state in inclusive and diffractive interactions is generated in a  $2 \rightarrow 2$ subprocess.  In contrast, in the exclusive production we have that the typical transverse momentum of the $\eta_c$ in the final state is determined by the transferred momentum  in the Odderon - proton vertex, which is larger than that present in the photon - proton vertex (See e.g. Refs. \cite{vicgustavo,vicdiego,motyka_odderon}). As the exclusive cross section has an $e^{ - \beta |t|}$ behavior, where $\beta$ is the slope parameter associated,  the associated $p_T$ distribution decreases exponentially at large transverse momentum. Therefore, it is expected that the production of $\eta_c$ with a large $p_T$ should be dominated by the inclusive and diffractive mechanisms.

\section{Summary}
\label{conc}

The description of the  mechanism underlying the production and decay of quarkonium still is  a theme of intense debate in literature. Significant theoretical improvements have been achieved in recent years and abundant experimental data have been accumulated at the LHC, in particular for the $J/\Psi$ production. The yield and $p_T$ distribution for the $J/\Psi$ production in hadronic collisions, dominated by gluon - gluon interactions, are quite well described by the NRQCD formalism. However, the description of the polarization of prompt $J/\Psi$ using this formalism still remains a theme of dispute. As a consequence, the study of $\eta_c$ provide an opportunity to further test the NRQCD formalism. Previous studies have focused in the $\eta_c$ production in hadronic colliders by gluon - gluon interactions. However, this particle also can be produced by photon - induced interactions. During the last years, the experimental results from Tevatron, RHIC and LHC have demonstrated that the study of  hadronic physics using photon - induced interactions in $pp/pA/AA$ colliders is feasible and provide important complementary information about the QCD dynamics and quarkonium production. In this paper we have complemented previous studies for the $\eta_c$ production by considering its production by photon - photon and inclusive and diffractive photon - hadron interactions. Our basic motivation to perform this study was the fact that the $\eta_c$ photoproduction is dominated by the color - octet process, which implies that this process provide an important test for the color - octet mechanism present in the NRQCD formalism. Moreover, in order to use the exclusive $\eta_c$ photoproduction as a direct probe of the Odderon, it is fundamental to known the magnitude of the $\gamma \gamma$ and diffractive $\gamma p$ channels, which also are characterized by two intact hadrons and two rapidity gaps in the final state. In this paper we have estimated the rapidity distributions for the $\eta_c$ production by $\gamma \gamma$, $\gamma h$ and $\gamma \pom$ interactions in $pp$ and $pPb$ collisions at the Run2 energies of the LHC. We have demonstrated that the inclusive $\eta_c$ photoproduction is not negligible and its study can be useful to test the NRQCD formalism. In the case of events with two rapidity gaps in the final state, we shown that the production of $\eta_c$ by $\gamma \pom$ interactions is dominant. Moreover, our results  indicated that the $\gamma \gamma$ channel also dominates the exclusive $\eta_c$ photoproduction. As a consequence, the probe of Odderon in exclusive processes is not an easy task, which will depend of the measurement of the transverse momentum of the hadrons in the final state and the implementation of a strict criteria of exclusivity in the events.


\section*{Acknowledgements}
This work was  partially financed by the Brazilian funding
agencies CNPq,  FAPERGS and  INCT-FNA (process number 
464898/2014-5).




\begin{thebibliography}{99}



\bibitem{review_nrqcd}
  N.~Brambilla, S.~Eidelman, B.~K.~Heltsley, R.~Vogt, G.~T.~Bodwin, E.~Eichten, A.~D.~Frawley and A.~B.~Meyer {\it et al.},
  Eur.\ Phys.\ J.\ C {\bf 71}, 1534 (2011)


\bibitem{nrqcd}
  G.~T.~Bodwin, E.~Braaten and G.~P.~Lepage,
  Phys.\ Rev.\ D {\bf 51}, 1125 (1995)
  [Erratum-ibid.\ D {\bf 55}, 5853 (1997)]


\bibitem{bk} 
  M.~Butenschoen and B.~A.~Kniehl,
  Phys.\ Rev.\ Lett.\  {\bf 104}, 072001 (2010)


\bibitem{bert}
  V.~P.~Goncalves and C.~A.~Bertulani,
  Phys.\ Rev.\ C {\bf 65}, 054905 (2002).
  

 \bibitem{vicmag}
  V.~P.~Goncalves and M.~V.~T.~Machado,
  Eur.\ Phys.\ J.\  C {\bf 40}, 519 (2005);   Phys.\ Rev.\  C {\bf 73}, 044902 (2006);
  Phys.\ Rev.\  D {\bf 77}, 014037 (2008);
  Phys.\ Rev.\ C {\bf 84}, 011902 (2011)

  \bibitem{brunoall} 
  V.~P.~Goncalves, B.~D.~Moreira and F.~S.~Navarra,
  Phys.\ Rev.\ C {\bf 90}, 015203 (2014);
 Phys.\ Lett.\ B {\bf 742}, 172 (2015);
  Phys.\ Rev.\ D {\bf 95}, no. 5, 054011 (2017)  
  

\bibitem{vicmairon} 
  V.~P.~Goncalves and M.~M.~Machado,
  Eur.\ Phys.\ J.\ A {\bf 50}, 72 (2014)

\bibitem{vicbruno_jpsidif}
V.~P.~Goncalves, L.~S.~Martins and B.~D.~Moreira,
  Phys.\ Rev.\ D {\bf 96}, no. 7, 074029 (2017)
  
  

\bibitem{upc}
 G. Baur, K. Hencken, D. Trautmann, S. Sadovsky, Y. Kharlov, Phys.
Rep. {\bf 364}, 359 (2002); 
V.~P.~Goncalves and M.~V.~T.~Machado,
Mod. Phys. Lett. A {\bf 19}, 2525  (2004); 
 C.~A. Bertulani, S.~R.~Klein and J.~Nystrand, Ann. Rev. Nucl. Part. Sci. {\bf 55}, 
271 (2005);
 K.~Hencken {\it et al.},
  Phys.\ Rept.\  {\bf 458}, 1 (2008). 



\bibitem{cdf} 
  T.~Aaltonen {\it et al.}  [CDF Collaboration],
  Phys.\ Rev.\ Lett.\  {\bf 102}, 242001 (2009)
  
\bibitem{star} 
  C.~Adler {\it et al.}  [STAR Collaboration],
  Phys.\ Rev.\ Lett.\  {\bf 89}, 272302 (2002)
  
  \bibitem{phenix} 
  S.~Afanasiev {\it et al.}  [PHENIX Collaboration],
  Phys.\ Lett.\ B {\bf 679}, 321 (2009)

\bibitem{alice} 
  B.~Abelev {\it et al.}  [ALICE Collaboration],
  Phys.\ Lett.\ B {\bf 718}, 1273 (2013)


\bibitem{alice2} 
  E.~Abbas {\it et al.}  [ALICE Collaboration],
  Eur.\ Phys.\ J.\ C {\bf 73}, 2617 (2013)
  
\bibitem{lhcb} 
  R. Aaij {\it et al.}  [LHCb Collaboration],
  J.\ Phys.\ G {\bf 40}, 045001 (2013)


\bibitem{lhcb2} 
  R. Aaij {\it et al.}  [LHCb Collaboration],
   J.\ Phys.\ G {\bf 41}, 055002 (2014)
   
\bibitem{lhcb3} 
  R.~Aaij {\it et al.} [LHCb Collaboration],
  JHEP {\bf 1509}, 084 (2015)
 
\bibitem{lhcbconf} 
  R.~Aaij {\it et al.} [LHCb Collaboration],
  LHCb-CONF-2016-007.
  
\bibitem{IS} G. Ingelman and P.E. Schlein, Phys. Lett. {\bf B152}, 256 (1985).


\bibitem{review_forward} 
  K.~Akiba {\it et al.} [LHC Forward Physics Working Group Collaboration],
  J.\ Phys.\ G {\bf 43}, 110201 (2016)


\bibitem{haoprl99} 
  L.~K.~Hao, F.~Yuan and K.~T.~Chao,
  Phys.\ Rev.\ Lett.\  {\bf 83}, 4490 (1999)
  
\bibitem{lhcb_etac}
  R.~Aaij {\it et al.} [LHCb Collaboration],
  Eur.\ Phys.\ J.\ C {\bf 75} (2015) no.7,  311  
  
  
\bibitem{kniel_prl15} 
  M.~Butenschoen, Z.~G.~He and B.~A.~Kniehl,
  Phys.\ Rev.\ Lett.\  {\bf 114}, no. 9, 092004 (2015)
  
\bibitem{han_prl15}
  H.~Han, Y.~Q.~Ma, C.~Meng, H.~S.~Shao and K.~T.~Chao,
  Phys.\ Rev.\ Lett.\  {\bf 114} (2015) no.9,  092005
  
\bibitem{zhang_prl15} 
  H.~F.~Zhang, Z.~Sun, W.~L.~Sang and R.~Li,
  Phys.\ Rev.\ Lett.\  {\bf 114}, no. 9, 092006 (2015)      
  
\bibitem{vic_odderon} 
  V.~P.~Goncalves,
  Nucl.\ Phys.\ A {\bf 902}, 32 (2013)
  
\bibitem{vic_odderon2} 
  V.~P.~Goncalves and W.~K.~Sauter,
  Phys.\ Rev.\ D {\bf 91}, no. 9, 094014 (2015)    
  
  
  
\bibitem{bkp} 
  J.~Bartels,
  Nucl.\ Phys.\ B {\bf 175}, 365 (1980); J.~Kwiecinski and M.~Praszalowicz,
  Phys.\ Lett.\ B {\bf 94}, 413 (1980).


\bibitem{ewerz} 
  C.~Ewerz,
  hep-ph/0306137.


  
\bibitem{martynov} 
  E.~Martynov and B.~Nicolescu,
  arXiv:1711.03288 [hep-ph].

\bibitem{khoze} 
  V.~A.~Khoze, A.~D.~Martin and M.~G.~Ryskin,
  arXiv:1801.07065 [hep-ph].



\bibitem{totem} 
  G.~Antchev {\it et al.} [TOTEM Collaboration],
  arXiv:1712.06153 [hep-ex]; 
  
  
\bibitem{totem2} 
  G. Antchev et al.. [TOTEM Collaboration]. CERN-EP-2017-335 
  
  
  
\bibitem{BF90} G. Baur, L.G. Ferreira Filho, Nucl. Phys. A {\bf 518}, 786 (1990). 	

\bibitem{Low}       F.~E.~Low,  Phys.\ Rev.\  {\bf 120}, 582 (1960). 				

\bibitem{h1} 
  F.~D.~Aaron {\it et al.}  [H1 Collaboration],
  Eur.\ Phys.\ J.\ C {\bf 68}, 401 (2010)

\bibitem{ko} 
  P.~Ko, J.~Lee and H.~S.~Song,
  Phys.\ Rev.\ D {\bf 54}, 4312 (1996)
  [Erratum-ibid.\ D {\bf 60}, 119902 (1999)]


\bibitem{cteq6}
J.~Pumplin, D.~R.~Stump, J.~Huston, H.~L.~Lai, P.~M.~Nadolsky and W.~K.~Tung,
  JHEP {\bf 0207}, 012 (2002)



\bibitem{H1diff} H1 Collab.,  A. Aktas {\it et al.}, Eur. Phys. J. {\bf C48} (2006) 715.


\bibitem{kluga}     M.~Klusek-Gawenda and A.~Szczurek, Phys.\ Rev.\ C {\bf 82}, 014904 (2010).	



\bibitem{Dress} M.~Drees and D.~Zeppenfeld, Phys.\ Rev.\ D {\bf
39}, 2536 (1989).



\bibitem{Kniehl}
B.~A.~Kniehl,
Phys.\ Lett.\ B {\bf 254}, 267 (1991).

\bibitem{frankfurt_ups} 
  L.~Frankfurt, V.~Guzey, M.~Strikman and M.~Zhalov,
  JHEP {\bf 0308}, 043 (2003); 
  V.~Guzey and M.~Zhalov, JHEP {\bf 1310}, 207 (2013); JHEP {\bf 1402}, 046 (2014).
 



\bibitem{Schafer_ups} 
  W.~Schafer and A.~Szczurek,
  Phys.\ Rev.\ D {\bf 76}, 094014 (2007);
 A.~Rybarska, W.~Schafer and A.~Szczurek,
  Phys.\ Lett.\ B {\bf 668}, 126 (2008);   A.~Cisek, W.~Schafer and A.~Szczurek,
  Phys.\ Rev.\ C {\bf 86}, 014905 (2012)
 
  
\bibitem{Martin} 
  S.~P.~Jones, A.~D.~Martin, M.~G.~Ryskin and T.~Teubner, JHEP {\bf 1311}, 085 (2013);  Eur.\ Phys.\ J.\ C {\bf 76}, no. 11, 633 (2016)


  
\bibitem{elements} 
  G.~M.~Yu, Y.~B.~Cai, Y.~D.~Li and J.~S.~Wang,
  Phys.\ Rev.\ C {\bf 95}, no. 1, 014905 (2017)
  Addendum: [Phys.\ Rev.\ C {\bf 95}, no. 6, 069901 (2017)]

  

\bibitem{vicdiego} 
  V.~P.~Goncalves, F.~S.~Navarra and D.~Spiering,
  Phys.\ Lett.\ B {\bf 768}, 299 (2017)  
  

\bibitem{vicgustavo} 
  G.~Gil da Silveira, V.~P.~Goncalves and M.~M.~Jaime,
  Phys.\ Rev.\ D {\bf 95}, no. 3, 034020 (2017)



  



\bibitem{motyka_odderon}   A.~Bzdak, L.~Motyka, L.~Szymanowski and J.~-R.~Cudell,
  Phys.\ Rev.\ D {\bf 75}, 094023 (2007)


 

\end{thebibliography}
\end{document}